# SCALING LAWS FOR ELECTROSTATIC COLLISIONLESS DRIFT INSTABILITIES

A.D. Bolshakova, A.Yu. Chirkov, V.I. Khvesyuk

*Bauman Moscow State Technical University, Moscow, Russia*
e-mail: alexxeich@mail.ru

Electrostatic drift instabilities driven by ion and electron temperature gradients (ITG and ETG instabilities) are considered for the ranges of perpendicular wave number values from the inverse ion thermal gyroradius up to the electron thermal gyroradius. The analysis is carried out in the framework of the local kinetic approach. Scaling laws for instability parameters are presented. These scaling laws can be used as a basis of collisionless plasma transport rates in magnetic traps with purely poloidal magnetic field topology.

## 1. Introduction

Systems with purely poloidal magnetic field topology have an important advantage (relative to the classical tokamak) which is the possibility of plasma confinement with $\beta \approx 1$ ($\beta$ = plasma pressure / magnetic pressure). The class of such systems includes open traps and cusps; field reversed configuration (FRC) combining open and closed magnetic field lines; closed mirror based systems; and traps with internal current-carrying conductors. The model of collisionless electromagnetic gradient drift instabilities was developed recently for such systems [1–3]. Studies have shown that the parameters of the drift instabilities depend on the nature of the magnetic field non-uniformity. The non-uniformity of the magnetic field is associated with a transverse gradient of magnetic induction and the curvature of the magnetic field lines. But, pure finite $\beta$ effect may be not very important in some regimes with high transversal non-uniformity of the magnetic field. Besides, maximal growth rates correspond to the case of a uniform field and low $\beta$, when electrostatic approximation is applicable. Here we study electrostatic case to obtain the scaling laws for maximum growth rates depending on the parameters of the plasma.

Two typical ranges of the transversal wave number $k_\perp$ are considered: i) $k_\perp \sim 1/\rho_{Ti}$ and ii) $k_\perp \sim 1/\rho_{Te}$, where $\rho_{Ti}$ and $\rho_{Te}$ are the ion and electron thermal gyroradiuses. The first range corresponds to the ion temperature gradient (ITG) driven instability and the instability due to the plasma density gradient. The second range corresponds to electron temperature gradient (ETG) instability. Oscillations with $k_\perp \gg 1/\rho_{Te}$ usually are not interesting from the viewpoint of plasma confinement and anomalous turbulent transport in magnetic fusion devices.

## 2. Drift instability model

Here we study ITG/ETG instability taking into account non-adiabatic responses of both ions and electrons in whole range of $k_\perp$ under the consideration. The analysis is carried out in the framework of the local electromagnetic kinetic approach. Low frequency ($\omega \ll \omega_{ci}$, $\omega_{ci}$ is the ion cyclotron frequency) drift instabilities are studied on the basis of the linearized Vlasov equation and quasineutrality condition. The dispersion equation is [4, 5]

$$1+\left[1-\frac{\omega_{*e}}{\omega}\left(1-\frac{3}{2}\eta_e\right)\right]\xi_e Z(\xi_e)\Gamma_0(b_e)-$$

$$\frac{\omega_{*e}}{\omega}\eta_e\xi_e\left[\xi_e+\xi_e^2 Z(\xi_e)\right]\Gamma_0(b_e)-\frac{\omega_{*e}}{\omega}\eta_e\xi_e Z(\xi_e)b_e\Gamma_0(b_e)-$$

$$\frac{\omega_{*e}}{\omega}\eta_e\xi_e Z(\xi_e)b_e[\Gamma_1(b_e)-\Gamma_0(b_e)]=-\tau\left\{1+\left[1+\frac{\omega_{*i}}{\omega}\left(1-\frac{3}{2}\eta_i\right)\right]\xi_i Z(\xi_i)\Gamma_0(b_i)+\right.$$

$$\left.\frac{\omega_{*i}}{\omega}\eta_i\xi_i\left[\xi_i+\xi_i^2 Z(\xi_i)\right]\Gamma_0(b_i)+\frac{\omega_{*i}}{\omega}\eta_i\xi_i Z(\xi_i)\Gamma_0(b_i)+\frac{\omega_{*i}}{\omega}\eta_i\xi_i Z(\xi_i)b_i[\Gamma_1(b_i)-\Gamma_0(b_i)]\right\}. \quad (1)$$

Here $\omega_{*j}=k_\perp\frac{k_B T_i}{q_j B L_n}$, $q_j$ is the charge of particle of kind "$j$" ($j = i, e$), $k_B$ is the Boltzmann constant, $B$ is the static magnetic field inside the plasma, $\omega$ is the complex frequency of the wave, $\tau = T_e/T_i$, $T_e$ is the electron temperature, $T_i$ is the ion temperature, $\eta_e = L_n/L_{Te}$, $\eta_i = L_n/L_{Ti}$, $L_{Te} = -T_e/\nabla_\perp T_e$, $L_{Ti} = -T_i/\nabla_\perp T_i$, $L_n = -n/\nabla_\perp n$, $n$ is the plasma density, $\Gamma_n(b)=I_n(b)\exp(-b)$, $I_n(b)$ is the modified Bessel function, $b_j = k_\perp^2 \rho_{Tj}^2$; $\rho_{Tj}=\frac{m_j v_{Tj}}{|q_j|B}$ is thermal gyroradius, $m_j$ is the mass of particle,

$$Z(\xi)=\frac{1}{\sqrt{\pi}}\int_{-\infty}^{\infty}\frac{e^{-u^2}du}{u-\xi} \quad (2)$$

is the plasma dispersion function of argument $\xi_j=\frac{\omega}{k_\parallel\sqrt{2k_B T_j/m_j}}$.

The difference in the directions of the diamagnetic drift of ions and electrons is taken into account in signs of terms of Eq. (1), where it is denoted $\omega_{*e}=k_\perp\frac{k_B T_e}{eBL_n}>0$.

Each solution of dispersion equation $\omega = \omega_R + i\gamma$ ($\omega_R$ is the real frequency, $\gamma$ is the growth rate) depends on the following parameters: parallel wave number $k_\parallel$, transversal wave number $k_\perp$, $\eta_i = L_n/L_{Ti}$, $\eta_e = L_n/L_{Te}$, $\tau = T_e/T_i$.

### 3. Results of the calculations

As the scale of growth rate $\gamma$ and real frequency $\omega_R$ we use $\omega_0 = k_B T_i/(eBL_n\rho_{Ti})$. Dimensionless wave numbers are $k_\parallel L_n$, and $k_\perp\rho_{Ti}$. In Figs. 1–5, results of the calculations of the maximal values of growth rates are presented. These data show maximal growth rates over all possible values of $k_\parallel$ and $k_\perp$. Fitting formulas (scaling laws) are suggested for the ITG and ETG regimes. Instability induced by density gradient (with no temperature gradients) was considered separately.

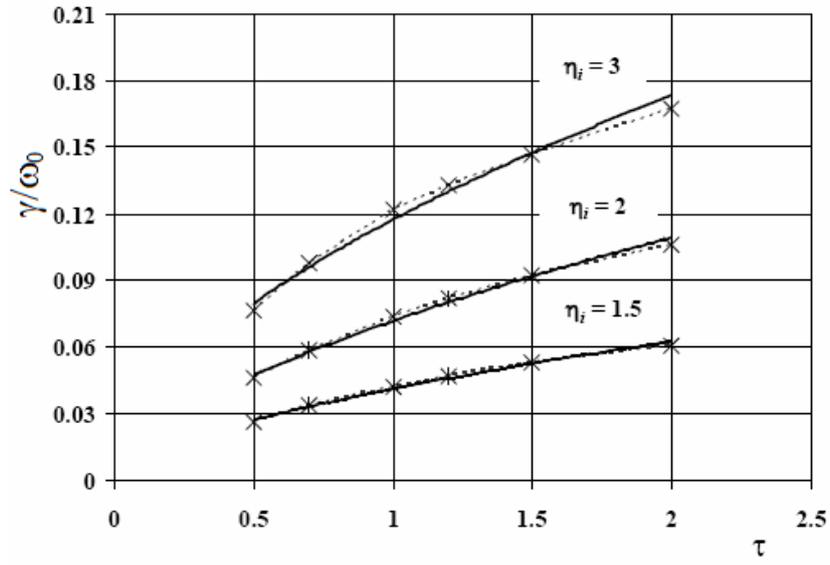

**Fig. 1.** Maximum growth rate vs $\tau$ at $k_\perp \rho_{Ti} = 1$ (ITG range), $\eta_i = 1.5$–$3$

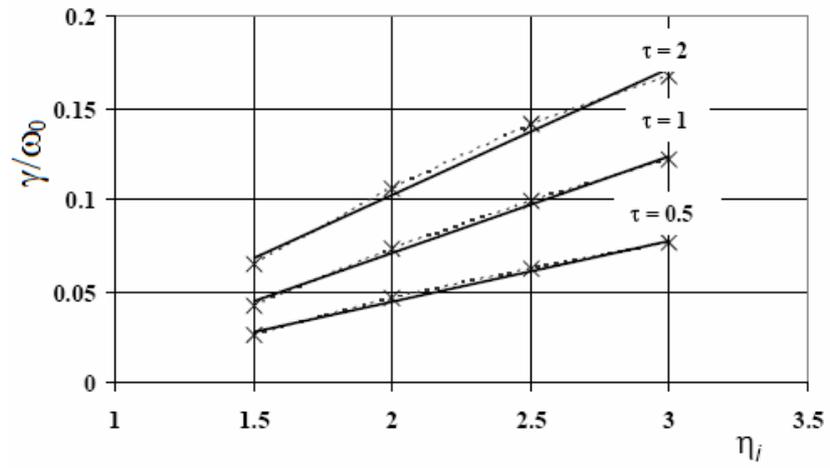

**Fig. 2.** Maximum growth rate vs $\eta_i$ at $k_\perp \rho_{Ti} = 1$ (ITG range), $\tau = 0.5$–$2$

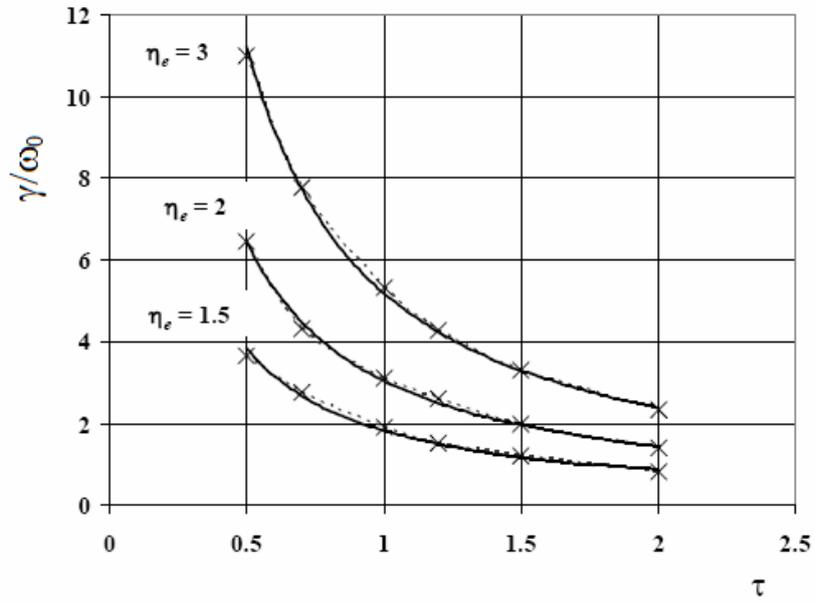

**Fig. 3.** Maximum growth rate vs $\tau$ at $k_\perp \rho_{Ti} = 40$ (ETG range), $\eta_e = 1.5$–3

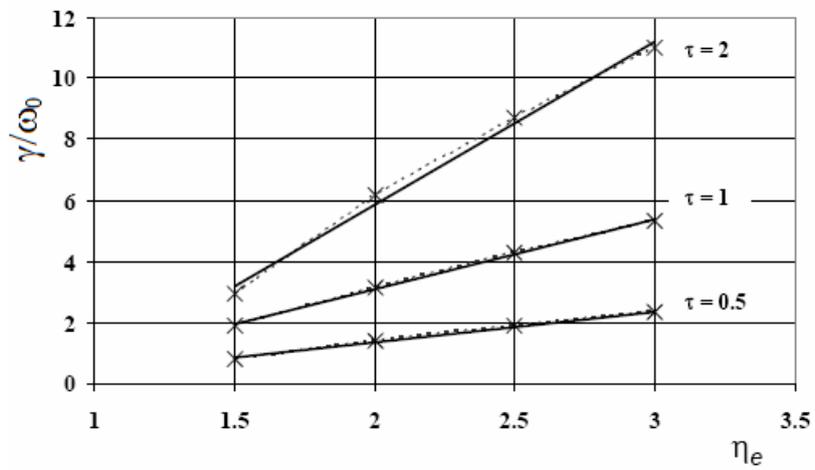

**Fig. 4.** Maximum growth rate vs $\eta_e$ at $k_\perp \rho_{Ti} = 40$ (ETG range), $\tau = 0.5$–2

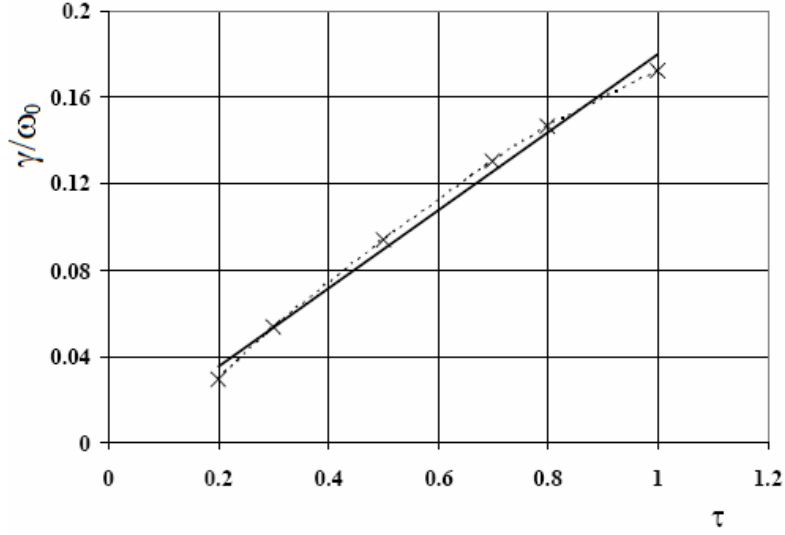

**Fig. 5.** Maximum growth rate vs $\tau$ at $k_\perp \rho_{Ti} = 0.5$, $\eta_i = \eta_e = 0$

For ITG range ($k_\perp \rho_{Ti} \sim 1$) at $\eta_i = 1.5\text{–}3$, and $\tau = 0.5\text{–}2$ scaling law is

$$\gamma_{max}/\omega_0 \approx 0.08 \tau^{0.6} (\eta_i - 0.8); \qquad (3)$$

for ETG range ($k_\perp \rho_{Te} \sim 1$) at $\eta_e = 1.5\text{–}3$, and $\tau = 0.5\text{–}2$

$$\gamma_{max}/\omega_0 \approx 4 \tau^{-1.1} (\eta_e - 0.8); \qquad (4)$$

at $\eta_i = \eta_e = 0$

$$\gamma_{max}/\omega_0 \approx 0.2\tau. \qquad (5)$$

These formulas can be used for transport rate estimations. For example, simple approximation for transversal diffusivity is

$$D_\perp \approx l^2 \gamma_{max}, \qquad (6)$$

where $l$ is the correlation length (typically $l \sim 1/k_\perp$ or $l \sim \rho_{Ti}$).

*Work was partially supported by RFBR grant 08-08-00459-a and Russian Presidential grant MK-1811.2010.8.*